\documentclass[aps,pra,preprint,superscriptaddress]{revtex4}
\usepackage{graphicx}
\usepackage{amsmath}

\begin{document}

\title{The second-order interference of two independent single-mode He-Ne lasers}

\author{Jianbin Liu}
\email[]{liujianbin@mail.xjtu.edu.cn}

\affiliation{Electronic Materials Research Laboratory, Key Laboratory of the Ministry of Education \& International Center for Dielectric Research, Xi'an Jiaotong University, Xi'an 710049, China}

\author{Minglan Le}
\affiliation{Electronic Materials Research Laboratory, Key Laboratory of the Ministry of Education \& International Center for Dielectric Research, Xi'an Jiaotong University, Xi'an 710049, China}

\author{Bin Bai}
\affiliation{Electronic Materials Research Laboratory, Key Laboratory of the Ministry of Education \& International Center for Dielectric Research, Xi'an Jiaotong University, Xi'an 710049, China}

\author{Wentao Wang}
\affiliation{Electronic Materials Research Laboratory, Key Laboratory of the Ministry of Education \& International Center for Dielectric Research, Xi'an Jiaotong University, Xi'an 710049, China}

\author{Hui Chen}
\affiliation{Electronic Materials Research Laboratory, Key Laboratory of the Ministry of Education \& International Center for Dielectric Research, Xi'an Jiaotong University, Xi'an 710049, China}

\author{Yu Zhou}
\affiliation{MOE Key Laboratory for Nonequilibrium Synthesis and Modulation of Condensed Matter, Department of Applied Physics, Xi'an Jiaotong University, Xi'an 710049, China}

\author{Fu-Li Li}
\affiliation{MOE Key Laboratory for Nonequilibrium Synthesis and Modulation of Condensed Matter, Department of Applied Physics, Xi'an Jiaotong University, Xi'an 710049, China}

\author{Zhuo Xu}
\affiliation{Electronic Materials Research Laboratory, Key Laboratory of the Ministry of Education \& International Center for Dielectric Research, Xi'an Jiaotong University, Xi'an 710049, China}

\begin{abstract}
The second-order spatial and temporal interference patterns with two independent single-mode He-Ne lasers are observed in a Hong-Ou-Mandel interferometer. Two-photon interference in Feynman's path integral theory is employed to interpret the experimental results. The conditions to observe the second-order interference pattern with two independent single-mode continuous wave lasers are discussed. It is concluded that two-photon interference exists for not only identical photons, but also photons with different spectrums if the detection system can not distinguish them in principle. The second-order temporal beating with two independent lasers can be employed to measure the coherence time and frequency of one laser if the properties of the other laser were known.
\end{abstract}


\date{\today}

\maketitle

\section{Introducntion}\label{introduction}

Interference is at the heart of quantum mechanics and ``\textit{it contains the only mystery}'' of quantum mechanics \cite{feynman-l}. Based on the conservation of energy, Dirac derived his famous statement about the first-order interference of light that ``\textit{Each photon then interferes only with itself. Interference between two different photons never occurs} \cite{dirac}.'' Soon after the invention of laser, the transient first-order interference pattern with two independent lasers was reported \cite{mandel-1963}. It seems that the observed interference pattern is due to the interference of photons emitted by different lasers, which contradicts Dirac's statement. Pfleegor and Mandel further proved that the transient first-order interference pattern exists when, with high probability, one photon is absorbed before the next one is emitted \cite{mandel-1967}. It is suggested by Paul that the first part of Dirac's statement is correct, while the second part is not always correct \cite{paul}. On the other hand, Mandel and Wolf suggest that the observed transient first-order interference pattern does not contradict Dirac's statement \cite{mandel-1965}. The detection of a photon forces the photon into a superposition state in which it is partly in each beam. ``It is the two components of the state of one photon which interfere, rather than two separate photons \cite{mandel-1965}.''

Mandel and Wolf further pointed out that the concept of photon is not helpful in understanding the transient first-order interference with two independent lasers \cite{mandel-1965}. However, this phenomenon can be well understood with the concept of photon if the superposition principle in Feynman's path integral theory is employed \cite{feynman-l,feynman-p}. When two independent lasers are superposed, there are two different ways to trigger a photon detection event in the observing plane. One way is the detected photon is emitted by one of the lasers and the other way the detected photon is emitted by the other laser. If these two different ways are indistinguishable, the probability amplitude of detecting a photon is the sum of these two probability amplitudes corresponding to the two different ways to trigger a photon detection event.  There is first-order interference. When these two different ways are distinguishable in principle, the probabilities instead of probability amplitudes should be added, in which there is no first-order interference. It is the probability amplitudes of different alternatives interfere, not different photons. Hence one can still observe the interference pattern when the intensity of two lasers are so low that, with high probability, there is only one photon in the system \cite{mandel-1967}.

Besides the studies on the transient first-order interference with two independent lasers, the second-order interference with two independent lasers has also been studied recently \cite{mandel-1967,jetp,ou-1988,ou-1989,kaltenbaek-2008,kaltenbaek-2009,liu-PRA,kim-2013,kim-2014}. All the interfering laser light beams are originated from the same laser except the ones in Refs. \cite{mandel-1967,jetp}. In Ref. \cite{mandel-1967}, they observed anticorrelation between single-photon counting rates of these two detectors and concluded that there is first-order interference pattern. As explained bellow, what they observed is not the second-order interference pattern, but a product of two first-order interference patterns measured by two detectors, respectively. In Ref. \cite{jetp}, they only measured the second-order spatial interference pattern in a Young's double-slit interferometer when the frequencies of these two lasers are close. In this paper, without any post-selection about the frequencies of the lasers, we will measure both the second-order spatial and temporal interference pattern with two independent single-mode He-Ne lasers in a Hong-Ou-Mandel (HOM) interferometer \cite{HOM}. The conditions to observe the second-order interference pattern with two independent single-mode continuous wave lasers are also discussed in Feynman's path integral theory. The studies are helpful to understand the physics of the second-order interference of light.

The rest parts of this paper are organized as follows. The theoretical calculations and experiments are in Sects. \ref{theory} and \ref{experiment}, respectively. Section \ref{discussion} includes the discussions based on the theoretical and experimental results. Our conclusions are in Sect. \ref{conclusion}.

\section{Theory}\label{theory}

There are two different theories to interpret the second-order interference of classical light. One is classical optical coherence theory based on statistical optics \cite{born,goodman-statistics}. The other one is Glauber's quantum optical coherence theory based on the wave mechanical formulation of quantum mechanics \cite{glauber-1}. Although the mathematical results in quantum and classical theories are equivalent for the interference of classical light \cite{sudarshan,glauber-1}, the physical interpretations are different in these two theories \cite{shih-book}. There are three different formulations of quantum mechanics, which are wave mechanics, matrix mechanics, and Feynman's path integral theory \cite{feynman-p,bohm,dirac}. Besides the optical coherence theory based on wave mechanical formulation \cite{glauber-1}, there should be optical coherence theory based on the matrix mechanical formulation and Feynman's path integral formulation, too. Comparing to the formulations of wave mechanics and matrix mechanics, Feynman's path integral formulation has the advantages of simplicity and easy to understand the physics of the calculations \cite{zee}. Hence we will employ two-photon interference theory based on Feynman's path integral formulation of quantum mechanics to interpret the second-order interference of two independent single-mode lasers. In fact, Feynman himself had employed path integral theory to discuss the first-order interference of light in a Young's double-slit interferometer \cite{feynman-l,feynman-p}, and two-photon bunching of thermal light \cite{feynman-q}. Fano employed Feynman's path integral theory to discuss the second-order interference of two photons emitted by two independent atoms, respectively \cite{fano}. In our earlier studies, we also employed this method to discuss the subwavelength interference \cite{liu-PRA}, the relationship between the first- and second- order interference patterns \cite{liu-OC}, and the second-order interference with laser and thermal light \cite{liu-OE,liu-EPL,liu-submitted}. This method indeed shows the advantage of understanding the interference of light better.

\begin{figure}[htb]
    \centering
    \includegraphics[width=80mm]{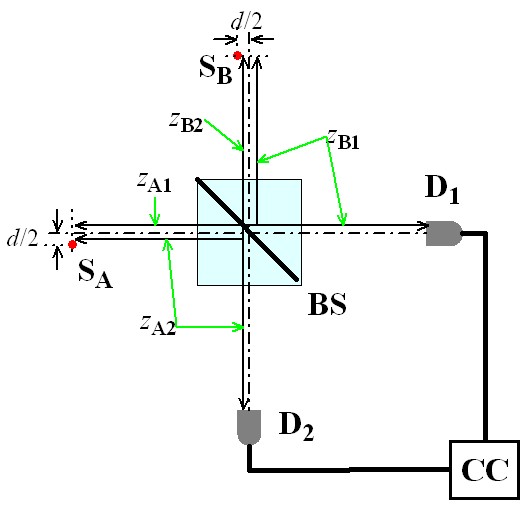}
    \caption{The HOM interferometer with two independent lasers. S$_A$ and S$_B$ are two independent single-mode laser point sources. BS is a $50:50$ nonpolarized beam splitter. D$_1$ and D$_2$ are two single photon detectors. CC is two-photon coincidence count system. $z_{A2}$ is the distance between S$_A$ and D$_2$ via BS. $z_{A1}$, $z_{B1}$, and $z_{B2}$ are defined similarly. The distances between the point sources and the symmetrical plane are equal, which is $d/2$.
    }\label{HOM-calculate}
\end{figure}

There are four different ways to trigger a two-photon coincidence count in Fig. \ref{HOM-calculate} \cite{liu-PRA,feynman-p}. One is these two photons are both emitted by S$_A$. The second one is these two photons are both emitted by S$_B$. The third one is the photon emitted by S$_A$ goes to D$_1$ and the photon emitted by S$_B$ goes to D$_2$. The last one is the photon emitted by S$_A$ goes to D$_2$ and the photon emitted by S$_B$ goes to D$_1$. The key to employ Feynman's path integral theory to calculate the two-photon probability is to judge whether these different alternatives are distinguishable or not. Feynman employes Heisenberg's uncertainty principle to discuss whether the alternatives to trigger a single-photon detection event in the Young's double-slit interferometer are distinguishable or not \cite{feynman-l,feynman-p}. We will follow Feynman's method to discuss whether these different alternatives to trigger a two-photon coincidence count are distinguishable or not.

In order to simplify the discussion, we assume the polarizations, intensities, and frequency bandwidths of these two lasers are the same. Photons are indistinguishable within one coherence volume \cite{mandel-1965,mandel-book}. When the time difference between these two single-photon detection events in a coincidence count is less than the coherence time of the laser, all these four different ways are indistinguishable. The two-photon probability distribution is  \cite{feynman-p,liu-PRA}
\begin{eqnarray}\label{G12-third}
&&G^{(2)}(\vec{r}_1,t_1;\vec{r}_2,t_2)\nonumber \\
&=& \langle |e^{i\varphi_A}K_{A1}e^{i(\varphi_A+\frac{\pi}{2})}K_{A2}+e^{i(\varphi_B+\frac{\pi}{2})}K_{B1}e^{i\varphi_B}K_{B2} \nonumber \\
&+&e^{i\varphi_A}K_{A1}e^{i\varphi_B}K_{B2} + e^{i(\varphi_A+\frac{\pi}{2})}K_{A2}e^{i(\varphi_B+\frac{\pi}{2})}K_{B1}|^2 \rangle
\end{eqnarray}
where we have written $K_\alpha(\vec{r}_j,t_j)$ as $K_{\alpha j}$ for short ($\alpha=A$, and $B$. $j=1$, and 2). $K_\alpha(\vec{r}_j,t_j)$ is Feynman's photon propagator that the photon emitted by Source $\alpha$ goes to D$_j$. $\varphi_{\alpha}$ is the initial phase of photon emitted by S$_\alpha$, which is a constant during the coherence time and fluctuates randomly between different coherence times \cite{handbook}. The extra phase $\pi/2$ is due to the photon reflected by a beam splitter will gain an extra phase comparing to the transmitted one \cite{loudon}$. \langle ... \rangle$ means ensemble average, which can be treated as time average in our experiment \cite{mandel-book}. When these two lasers are independent, $\langle e^{i(\varphi_A-\varphi_B)}\rangle $ equals 0. Equation (\ref{G12-third}) can be simplified as
\begin{eqnarray}\label{G12-fourth}
&&G^{(2)}(\vec{r}_1,t_1;\vec{r}_2,t_2)\nonumber\\
&=& |K_{A1}K_{A2}|^2+|K_{B1}K_{B2}|^2+|K_{A1}K_{B2}|^2+|K_{A2}K_{B1}|^2 \nonumber \\
&-& (K^*_{A1}K^*_{B2}K_{A2}K_{B1}+ c.c.),
\end{eqnarray}
where $c.c.$ means complex conjugation and the minus sign on the righthand side of Eq. (\ref{G12-fourth}) is due the $\pi$ phase difference between the third and fourth terms in Eq. (\ref{G12-third}). The last two terms on the righthand side of Eq. (\ref{G12-fourth}) is due to the interference of the last two alternatives, with which the second-order interference pattern can be observed. The first two terms on the righthand side of Eq. (\ref{G12-third}) only contribute a constant background to the second-order interference pattern, which limits the visibility of second-order interference pattern \cite{mandel-1965}.

In the same condition, the one-photon probability distribution at D$_j$ ($j=1$, and 2) is
\begin{eqnarray}\label{G1-first}
&&G^{(1)}(\vec{r}_j)\nonumber\\
&=&\langle |e^{i\varphi_A}K_{Aj}+e^{i(\varphi_B+\frac{\pi}{2})}K_{Bj}|^2 \rangle \\
&=& \langle |K_{Aj}|^2 \rangle + \langle |K_{Bj}|^2 \rangle +[\langle e^{i(\varphi_A-\varphi_B-\frac{\pi}{2})}K_{Aj}K_{Bj}^* \rangle+ c.c.]. \nonumber
\end{eqnarray}
When these two lasers are independent, Eq. (\ref{G1-first}) can be simplified as
\begin{eqnarray}\label{G1-second}
&&G^{(1)}(\vec{r}_j)=\langle |K_{Aj}|^2 \rangle + \langle |K_{Bj}|^2 \rangle
\end{eqnarray}
Comparing Eqs. (\ref{G12-fourth}) and (\ref{G1-second}), although the first-order interference pattern can not be observed for long collecting time with two independent lasers,  the second-order interference pattern can be observed in the same condition.

When the collecting time is shorter than the coherence time, the first-order interference pattern is given by
\begin{eqnarray}\label{G1-new-1}
&&G^{(1)}(\vec{r}_j)\nonumber\\
&=&  |K_{Aj}|^2 + |K_{Bj}|^2  +[e^{i(\varphi_A-\varphi_B-\frac{\pi}{2})}K_{Aj}K_{Bj}^* + c.c.].
\end{eqnarray}
The first-order interference pattern can be observed for the initial phases of photons within the coherence time are constant \cite{mandel-1963,mandel-1967}. The second-order interference pattern can not be observed when the collecting time is shorter than the coherence time \cite{liu-OC}.

The second-order interference pattern in Eq. (\ref{G12-fourth}) can be further simplified by taking Feynman's photon propagator into consideration. For a point single-mode laser source, Feynman's photon propagator is \cite{peskin, liu-PRA}
\begin{eqnarray}\label{propagator}
K_{\alpha \beta}=\frac{\text{exp}[-i(\vec{k}_{\alpha\beta}\cdot
\vec{r}_{\alpha\beta}-\omega_{\alpha}
t_{\beta})]}{r_{\alpha\beta}},
\end{eqnarray}
which is the same as the Green function for a point source in classical optics \cite{born}. $\vec{k}_{\alpha\beta}$ and $\vec{r}_{\alpha\beta}$ are the wave and position vectors of the photon emitted by S$_\alpha$ and detected at D$_\beta$, respectively. $r_{\alpha\beta}=|\vec{r}_{\alpha\beta}|$ is the
distance between S$_\alpha$ and D$_\beta$. $\omega_{\alpha}$ and $t_{\beta}$ are the frequency and time for the photon that is emitted by S$_\alpha$ and detected at D$_\beta$, respectively. Substituting Eq. (\ref{propagator}) into Eq. (\ref{G12-fourth}), it is straightforward to have
\begin{eqnarray}\label{G12-five}
&&G^{(2)}(\vec{r}_1,t_1;\vec{r}_2,t_2)\nonumber\\
&=&\frac{4}{r^4}\{1-\frac{1}{2}\cos [(\vec{k}_{A1} \cdot
\vec{r}_{A1}-\vec{k}_{B1}\cdot \vec{r}_{B1})\nonumber \\
&&-(\vec{k}_{A2} \cdot \vec{r}_{A2}-\vec{k}_{B2}\cdot
\vec{r}_{B2})]\} \cos[\Delta \omega_{AB}(t_1-t_2)],
\end{eqnarray}
where the approximation $r_{\alpha\beta}\approx r_\beta \sim r$
($\alpha=A$, and $B$. $\beta=1$, and 2) has been employed. $r_\beta$ is the distance between the symmetrical position in the source plane and D$_\beta$. This approximation is valid when the distance $d$ is much smaller than the distance $L$ between the source and detection planes \cite{born}. $\Delta \omega_{AB}$ is the difference between the mean frequencies of these two He-Ne lasers. Assuming $|\vec{k}_{\alpha \beta}| \simeq k$ ($\alpha=$A and B, $\beta$=1 and 2) and only considering the one dimension case, the second-order coherence function in Eq. (\ref{G12-five}) can be further simplified as
\begin{eqnarray}\label{G12-six}
&&G^{(2)}(\vec{r}_1,t_1;\vec{r}_2,t_2)\nonumber\\
&=&\frac{4}{r^4}\{1-\frac{1}{2}\cos [\frac{kd}{L}(x_1-x_2)] \cos[\Delta \omega_{AB}(t_1-t_2)]\}.
\end{eqnarray}
Both the spatial and temporal second-order interference patterns can be observed in Eq. (\ref{G12-six}). The visibility of the second-order interference pattern is 50\%. Note that Eq. (\ref{G12-six}) is valid on condition that the time difference between these two photon detection events is shorter than the coherence time of the laser. When the time difference is longer than the coherence time, there will be no two-photon interference.

\section{Experiment}\label{experiment}

\begin{figure}[htb]
    \centering
    \includegraphics[width=80mm]{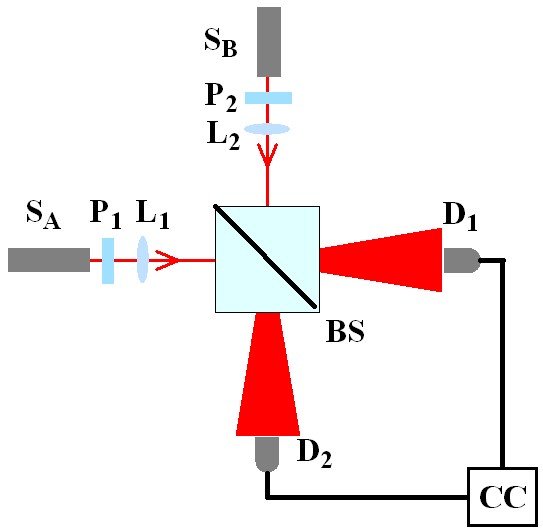}
    \caption{The experimental setup. S$_A$ and S$_B$ are two independent single-mode He-Ne lasers. P$_1$ and P$_2$ are two vertically polarized polarizers. L$_1$ and L$_2$ are two identical lens with focus lengths of 50 mm. BS is a $50:50$ nonpolarized beam splitter. D$_1$ and D$_2$ are two single photon detectors. CC is two-photon coincidence count system.
    }\label{setup}
\end{figure}

The experimental setup is shown in Fig. \ref{setup}, which is similar as the one in Fig. \ref{HOM-calculate}. Two point sources are simulated by two independent He-Ne lasers (DH-HN250P) focused by two identical lens. The polarizers P$_1$ and P$_2$ are employed to ensure that the polarizations of the light emitted by these two lasers are the same. The distance between the lens and the detector planes all equal 695 mm. The single photon detectors (SPCM-AQRH-14-FC) and two-photon coincidence counting system (SPC-630) are the same as the ones we employed before \cite{liu-PRA,liu-EPL,liu-OE,liu-OC,liu-submitted}.

\begin{figure}[htb]
    \centering
    \includegraphics[width=80mm]{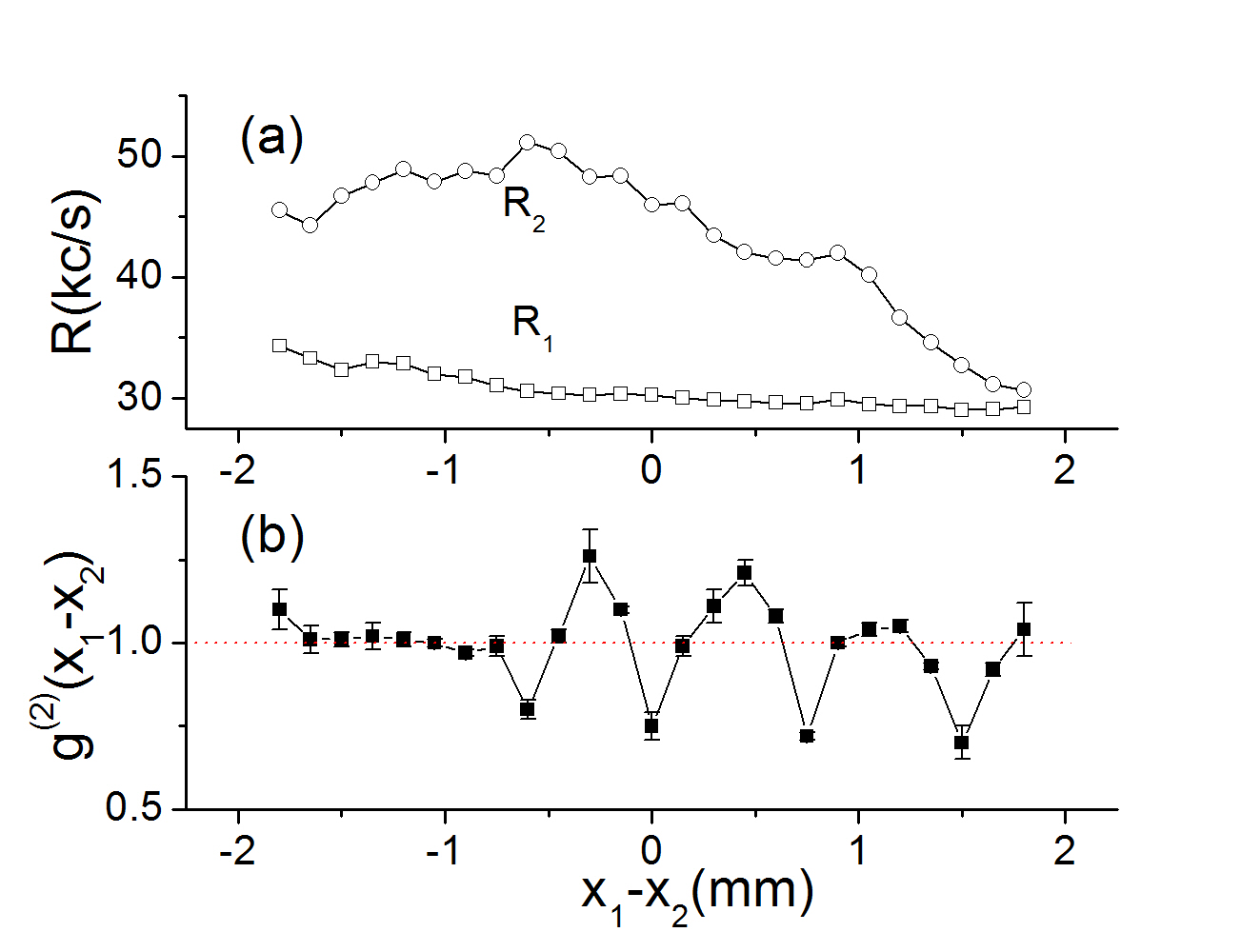}
    \caption{The second-order spatial interference pattern. R$_1$ and R$_2$ in (a) are the single-photon counting rates of D$_1$ and D$_2$, respectively. The normalized second-order coherence function is in (b). See text for detail.
    }\label{spatial}
\end{figure}

\begin{figure}[htb]
    \centering
    \includegraphics[width=80mm]{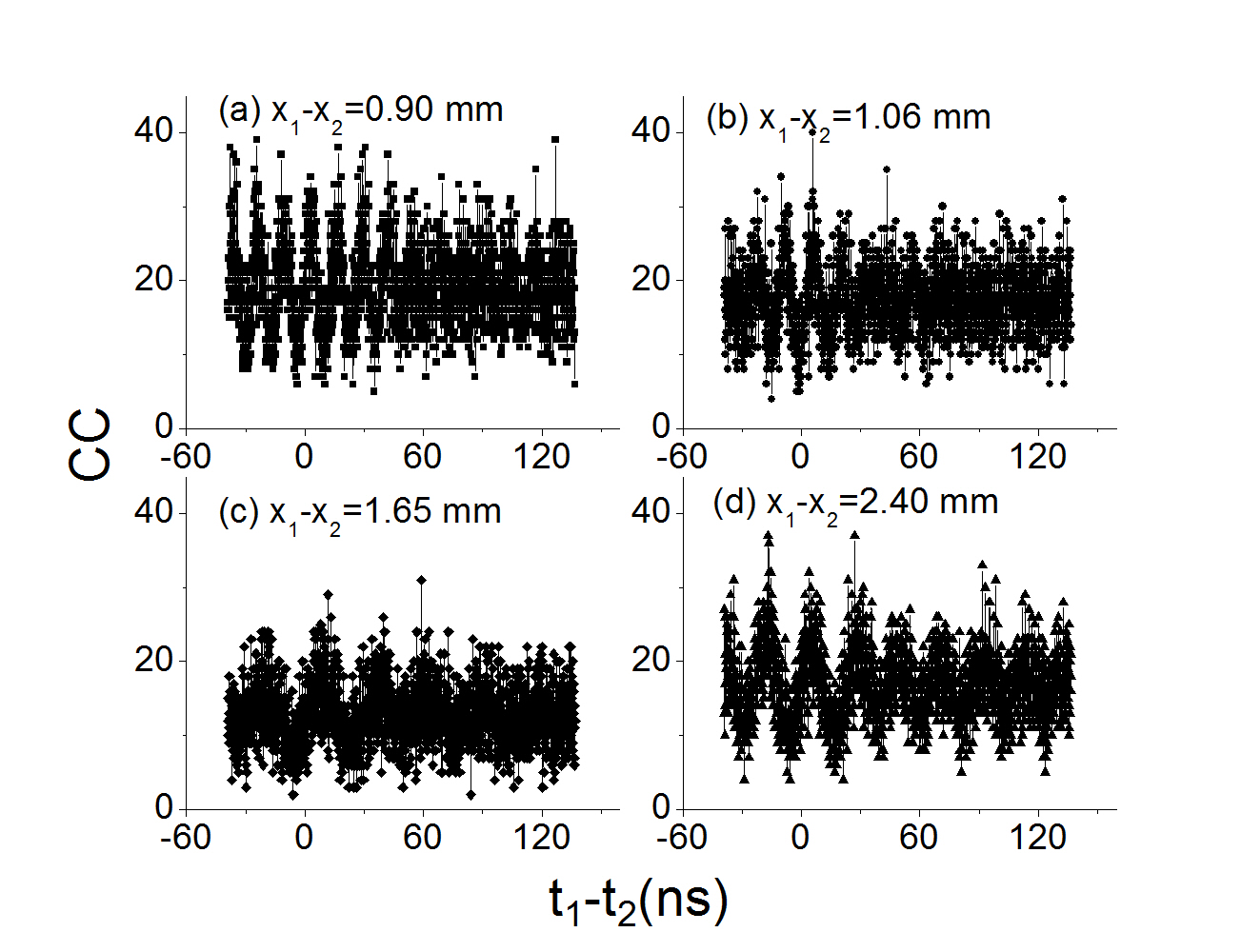}
    \caption{The second-order temporal interference pattern. These four figures are measured for different separations of the detectors. See text for details.
    }\label{temporal}
\end{figure}

We first measure the second-order spatial interference pattern by scanning the position of D$_2$ horizontally while keeping the position of $D_1$ fixed. The two-photon coincidence count time window in our experiment is 4.88 ns. The collecting time for each run is 120 s and at least three groups of data are collected for one position of D$_2$. The experimental results are shown in Fig. \ref{spatial}, where the single-photon counting rates of D$_1$ and D$_2$ are in Fig. \ref{spatial}(a) and the normalized second-order coherence functions are in Fig. \ref{spatial}(b). There is no first-order interference pattern as shown by the single-photon counting rates of D$_2$. The counting rates of D$_1$ are nearly constant, for it is fixed at the same position in the whole measurement. The decrease of R$_1$ is due to the energy of laser decreases with time. The second-order coherence function is shown in Fig. \ref{spatial}(b), where we have normalized it to the coherence function of single-mode laser \cite{glauber-1}. When the value of  $x_1-x_2$ is less than -0.5, the normalized second-order coherence function equals 1. These two single-photon detection events are independent \cite{liu-EPL}. The main reason is due to the frequency difference between these two lasers changes fast during the measurement. When $x_1-x_2$ is in the regime of [-0.5, 2], there is second-order spatial interference pattern. The observed spatial interference pattern is not strictly a cosine function as predicted by Eq. (\ref{G12-six}). The reason is also the frequency difference between these two lasers are not well fixed during the measurement, which can be seen from the measured second-order temporal beating in Fig. \ref{temporal}.

The typical measured second-order temporal interference patterns are shown in Fig. \ref{temporal}, in which no background is subtracted. These four different groups of data are collected when the position separations between these two detectors are 0.90 mm, 1.06 mm, 1.65 mm, and 2.40 mm in Figs. \ref{temporal}(a)-(d), respectively. Based on the second-order temporal beating, we can calculate the frequency bandwidth of these two lasers and the frequency difference between them. As discussed above, beating can only be observed within the coherence time of the laser. The measured coherence times in these four figures are 48.2 ns, 26.1 ns, 73.3 ns, and 122.4 ns, respectively. The beating periods in Figs. \ref{temporal}(a)-(d) are 7.0 ns, 13.8 ns, 28.4 ns, and 21.9 ns, respectively. The coherence time and beating period change for different measurements. In fact, they also change within one measurement \cite{handbook}. Only if the frequency changes are not so fast and large, we can observe the second-order interference pattern with two independent lasers as shown in Figs. \ref{spatial} and \ref{temporal}.

\section{Discussions}\label{discussion}

\begin{figure}[htb]
    \centering
    \includegraphics[width=80mm]{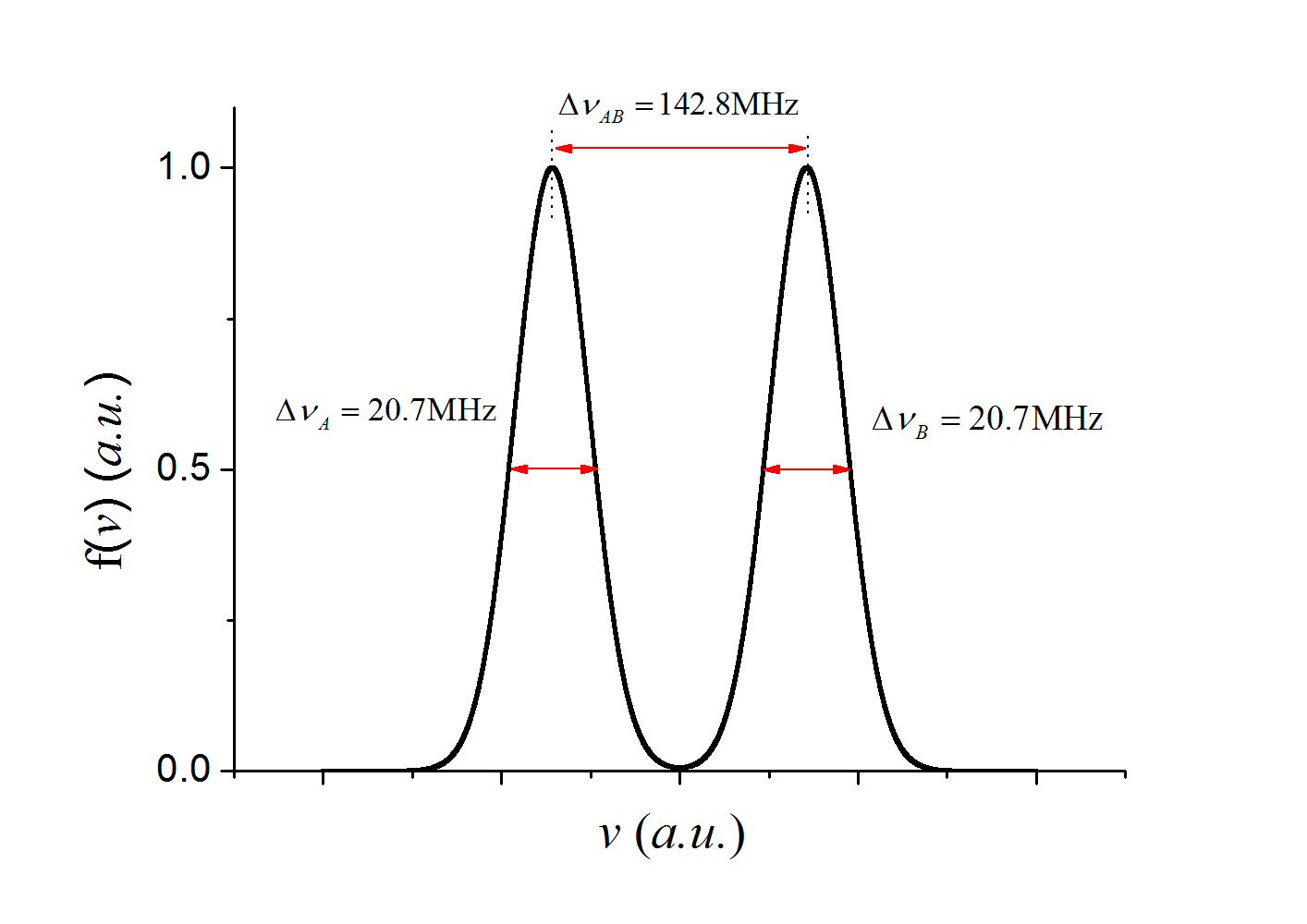}
    \caption{The frequency distributions of two He-Ne lasers. $\Delta \nu_{AB}$ is the frequency difference between these two lasers.  $\Delta \nu_{A}$ and $\Delta \nu_{B}$ are the frequency bandwidths of laser A and B, respectively. See text for details.
    }\label{separation}
\end{figure}

In the sections above, we have theoretically proved that there is second-order interference pattern by superposing two independent single-mode continuous wave lasers and experimentally verified it with two identical single-mode He-Ne lasers in a HOM interferometer. Although our experimental results can be explained in classical theory \cite{sudarshan,glauber-1}, interesting point can be made if quantum theory is employed. We will take the second-order temporal beating in Fig. \ref{temporal}(a) for example. The measured coherence time and beating period are 48.2 ns and 7.0 ns, respectively. The frequency bandwidth of these two lasers are both 20.7 MHz \cite{frequency}. The frequency difference between these two lasers is 142.8 MHz. Assuming the frequency distribution of single-mode He-Ne laser is Gaussian, the frequency distribution corresponding to the temporal beating in Fig. \ref{temporal}(a) is shown in Fig. \ref{separation}. The frequency separation is larger than the frequency bandwidths of these two lasers. The photons emitted by these two lasers are distinguishable by measuring their frequencies.

\begin{figure}[htb]
    \centering
    \includegraphics[width=50mm]{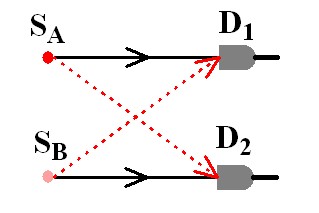}
    \caption{Two-photon interference. S$_A$ and S$_B$ are two independent light sources. See text for details.
    }\label{photon}
\end{figure}

In the case that each laser emits one photon to trigger a two-photon coincidence count in Fig. \ref{setup}, there are two different ways for these two photons to trigger a two-photon coincidence event as shown in Fig. \ref{photon}. One is photon A emitted by S$_A$ goes to D$_1$ and photon B emitted by S$_B$ goes to D$_2$. The other one is photon A goes to D$_2$ and photon B goes to D$_1$. When photons A and B are distinguishable, these two different ways are distinguishable since we know which photon is detected by which detector. There should be no second-order interference in the condition in Fig. \ref{temporal}(a). However, the second-order interference pattern is observed in this condition, which means there is two-photon interference. How can distinguishable alternatives interfere with each other? Does it not contradict the superposition principle in Feynman's path integral theory? The reason why there is two-photon interference with distinguishable photons is that the measurement in quantum mechanics is depend on the measuring apparatus. Although the photons emitted by these two lasers have different spectrums shown in Fig. \ref{separation}, these photons are indistinguishable for the detection system if the system can not distinguish them in principle. The Heisenberg's uncertainty principe is $\Delta \nu \Delta t > 1$ when considering the frequency and time measurement of a photon \cite{bohm}, where $\Delta \nu$ and $\Delta t$ are the measurement uncertainty of frequency and time, respectively. When the uncertainty of time measurement is $\Delta t$ for a detection system, this system can not distinguish photons with frequency difference less than $1/\Delta \nu$ in principle \cite{feynman-p,bohm}. If the time measurement uncertainty of a detection system is zero, the frequency measurement uncertainty is infinity. All the photons are indistinguishable for the detection system, which means there are two-photon interference for photons of different colors. However, it is impossible to have a detection system with zero time measurement uncertainty. The response time of our detection system is about 0.45 ns \cite{shih-book,thesis}, which can be treated as the uncertainty of time measurement. The detection system can not distinguish photons with frequency difference less than 2.2 GHz. The frequency difference of photons emitted by these two He-Ne lasers in Fig. \ref{temporal}(a) is 142.8 MHz, which is much less than 2.2 GHz. Hence these two photons are indistinguishable for our detection system and two-photon interference does happen.

In a recent paper, Kim \textit{et al.} discussed the condition for two-photon interference with coherent pulses. They claimed that coherence within the same input is essential for two-photon interference \cite{kim-2013}. Based on the superposition principle in Feynman's path integral theory, the necessary and sufficient condition for two-photon interference is there are more than one different yet indistinguishable alternatives to trigger a two-photon coincidence count event \cite{feynman-p}. What they found is the condition to observe the second-order interference pattern in their scheme \cite{kim-2013}. The condition to observe second-order interference pattern is different from the condition to have two-photon interference. If there is no two-photon interference, there is no second-order interference pattern. However, the second-order interference pattern may not be observable even if there is two-photon interference. Some conditions must be satisfied in order to observe the second-order interference pattern.

In the second-order interference of two independent single-mode continuous wave lasers, the conditions to observe the second-order interference pattern are as follows: (I) the photons emitted by these two lasers are indistinguishable for the detection system. (II) The phase difference between these two lasers changes randomly or changes in the way to erase the first-order interference pattern between these two lasers. (III) The central frequencies of these two lasers remain nearly constant during one measurement. The first condition requires the frequency difference between these two lasers is less than the uncertainty of  frequency measurement of the detection system. The second condition is automatically satisfied for two independent lasers as long as the collection time is much longer than the coherence time. If the relative phase between these two independent lasers is constant, there is no second-order interference pattern. The observed pattern is the product of two first-order interference patterns. The normalized second-order coherence function equals to 1 in this condition. This is also the reason why there is no second-order interference pattern when a single-mode laser is incident to a Young's double-slit interferometer or Michelson interferometer. Due to the same reason, there is no second-order interference pattern for two independent laser within the coherence time, in which the transient first-order interference pattern is observed \cite{mandel-1963,mandel-1967}. The observed anticorrelation in Pfleegor \textit{et al.}'s experiment is due to the product of two first-order interference patterns \cite{mandel-1967}. The normalized second-order coherence function in their experiment equals 1, which means these two single-photon detection events are independent \cite{liu-EPL}.

The method above can also be employed to discuss the second-order interference with nonclassical light \cite{mandel-1965}. For instance, if these two point sources are single-photon sources, the two-photon probability distribution is
\begin{eqnarray}\label{G12-photon}
G^{(2)}(\vec{r}_1,t_1;\vec{r}_2,t_2)=|e^{i\varphi_A}K_{A1}e^{i\varphi_B}K_{B2}+e^{i(\varphi_A+\frac{\pi}{2})}K_{A2}e^{i(\varphi_B+\frac{\pi}{2})}K_{B1}|^2,
\end{eqnarray}
when these two single photons are indistinguishable. Note that these two photons are not necessary to be identical in order to have two-photon interference. The reason why there are only two terms on the righthand side of Eq. (\ref{G12-photon}) instead of four as Eq. (\ref{G12-third}) is due to single-photon source can emit only one photon at a time. The initial phases of the emitted photons should be random in order to observe the second-order interference pattern. Equation (\ref{G12-photon}) can be simplified as
\begin{eqnarray}\label{G12-photon-s}
G^{(2)}(\vec{r}_1,t_1;\vec{r}_2,t_2)\propto 1-\cos [\frac{kd}{L}(x_1-x_2)] \cos[\Delta \omega_{AB}(t_1-t_2)],
\end{eqnarray}
in which all the approximations as the one in Eq. (\ref{G12-six}) have been employed. The visibility of the second-order interference pattern with single-photon sources is 100\%, which is consistent with the conclusion in Ref. \cite{mandel-1983}.

Based on the discussions for the second-order interference with classical and quantum light above, the conditions to have two-photon interference are the same for classical and quantum light. There is no difference in interference for these two kinds of light. The reason why the interference patterns are different for classical and quantum light is due to the properties of the light sources are different. Hence it is questionable to say classical light has classical interference while nonclassical light has quantum interference . There is indeed some difference between the classical and quantum interference as pointed out by Dirac \cite{dirac}. For instance, in the case of a classical system for which a superposition principle holds, a state superposed with itself will get a different state. While in quantum system, a state superposed with itself will get the same state. The difference between the superposition principles in classical and quantum theories is different from the one in the second-order interference with quantum and classical light.

\section{Conclusions}\label{conclusion}

In conclusion, we have observed the second-order spatial and temporal interference patterns with two independent single-mode He-Ne lasers in a HOM interferometer. The observed second-order temporal beating can be employed to measure the coherence time and frequency of one laser if the properties of the other laser were known. The necessary and sufficient condition to have two-photon interference is there are more than one different yet indistinguishable alternatives to trigger a two-photon coincidence count event. This conclusion can be generalized to the first-, third-, and high-order interference. We also discussed the conditions to observe the second-order interference pattern with two independent single-mode continuous wave lasers. Photons with different spectrums can have two-photon interference on condition that they are indistinguishable for the detection system. The interference with classical and quantum light are the same. The reason why interference patterns are different for these two kinds of light is due to the different properties of the light sources, not the interference itself.

\section*{Acknowledgement}
This project is supported by National Science Foundation of China (No. 11404255), Doctoral Fund of Ministry of Education of China (No. 20130201120013), the 111 Project of China (No. B14040), and the Fundamental Research Funds for the Central Universities.


\begin{thebibliography}{99}
\bibitem{frequency} We have assumed the frequency bandwidths of these two lasers are the same in the calculation. If the bandwidths of these two lasers are not the same, what we have measured is the coherence time of the laser with larger frequency bandwidth. The frequency bandwidths of these two lasers are less than or equal to 20.7 MHz.
\end{thebibliography}
\end{document}